\documentclass[twocolumn,showpacs,amsmath,amssymb,prl]{revtex4-1}
\usepackage{mathrsfs}
\usepackage{graphicx}
\usepackage{dcolumn}
\usepackage{bm}
\usepackage{amsmath}
\usepackage{amsfonts}

\begin{document}

\title{Quantum Anomalous Hall Effect in Magnetic Insulator Heterostructure}

\author{Gang Xu$^{1,2}$, Jing Wang$^1$, Claudia Felser$^3$, Xiao-Liang Qi$^1$, and Shou-Cheng Zhang$^1$}

\affiliation{$^1$ Department of Physics, McCullough Building, Stanford University, Stanford, CA 94305-4045, USA\\
$^2$ Beijing National Laboratory for Condensed Matter Physics, and Institute of Physics, Chinese Academy of Sciences, Beijing 100190, China\\
$^3$ Max Planck Institute for Chemical Physics of Solids, Dresden, Germany}

\date{\today}

\begin{abstract}
Based on \emph{ab initio} calculations, we predict that a monolayer of Cr-doped (Bi,Sb)$_2$Te$_3$ and GdI$_2$ heterostructure is a quantum anomalous Hall insulator with a non-trivial band gap up to 38~meV. The principle behind our prediction is that the band inversion between two topologically trivial ferromagnetic insulators can result in a non-zero Chern number, which offers a better way to realize the quantum anomalous Hall state without random magnetic doping. In addition, a simple effective model is presented to describe the basic mechanism of spin polarized band
inversion in this system. Moreover, we predict that 3D quantum anomalous Hall insulator could be realized in (Bi$_{2/3}$Cr$_{1/3} $)$_2$Te$_3$ /GdI$_2$ superlattice.
\end{abstract}

\maketitle

The recent discovery of quantum anomalous Hall (QAH) effect has attracted tremendous interest in condensed matter physics~\cite{Qi2006,Qi2008,Liu2008,Li2010,Yu2010,Xu2011,xiao2011,ruegg2011,Chang2013,wang2013a,wang2013b,zhang2014,Garrity2014,wang2014,kou2014,checkelsky2014}. In a
QAH insulator, theoretically predicted in magnetic topological insulators (TIs)~\cite{Qi2006,Qi2008,Liu2008,Li2010,Yu2010}, the strong spin-orbit coupling and ferromagnetic (FM) ordering combine to give rise to an insulating state with a topologically nontrivial band structure
characterized by a finite Chern number~\cite{Thouless1982,Haldane1988}. Recently, the QAH effect has been experimentally observed in Cr-doped (Bi,Sb)$_2$Te$_3$ around 30~mK~\cite{Chang2013}. The robust dissipationless chiral edge states in the QAH state could be used for interconnects of semiconductor devices. Unfortunately, the topologically nontrivial band gap of this system is extremely small and the quantization of Hall conductance can only be observed below about 100~mK~\cite{kou2014,checkelsky2014}. For potential device applications, it is important to
find materials for the QAH effect, and to increase the band gap as well as the Curie temperature
($T_c$) of magnetic moments.

\begin{figure}[t]
\includegraphics[scale=0.5,angle=270]{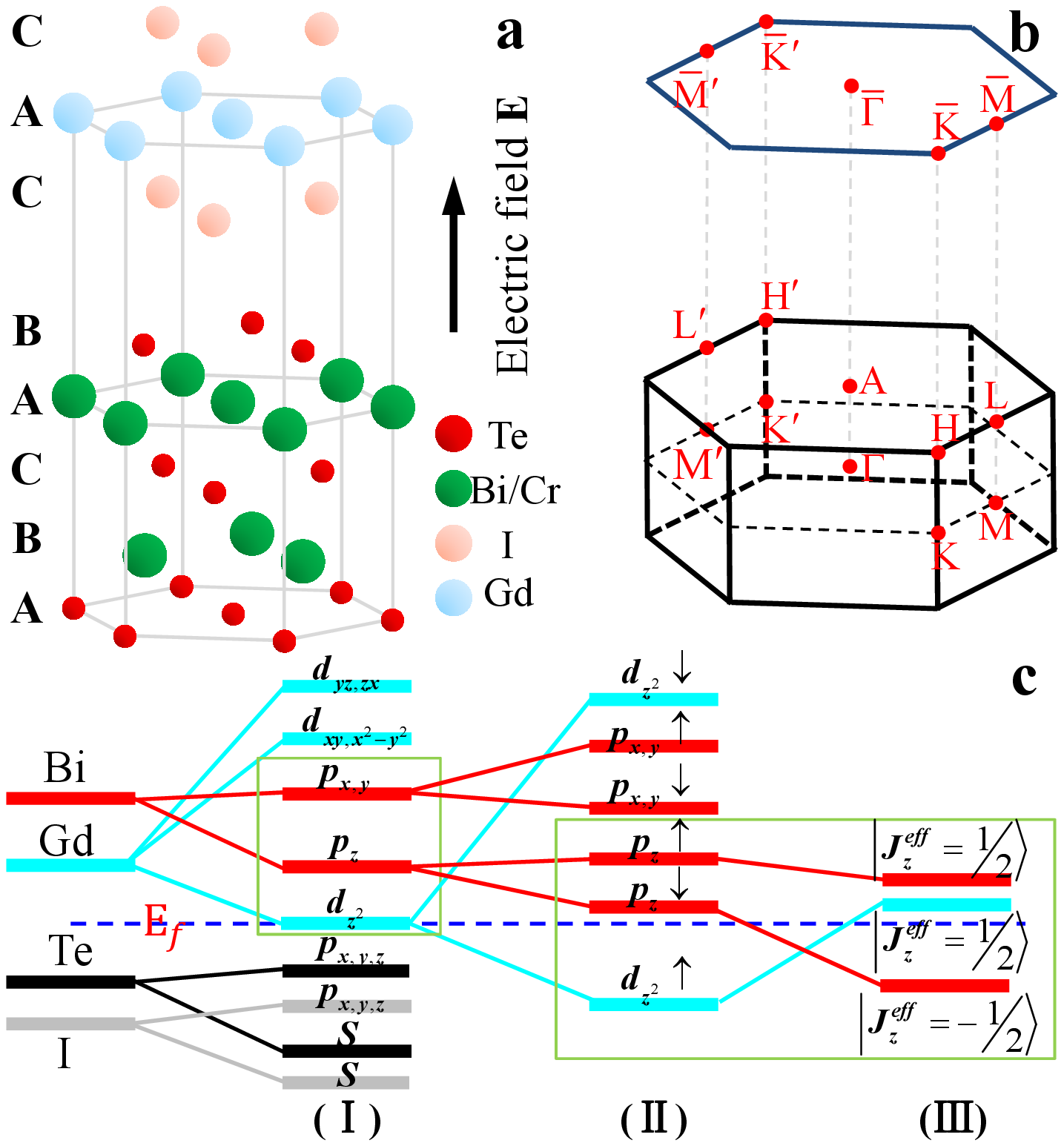}
\caption{(Color online) a, The crystal structure of monolayer Cr-doped Bi$_2$Te$_3$ and GdI$_2$ film. The black arrow indicate the electric field. b, The Brillouin zones (BZ) for the film (2D) and superlattice (3D) of triangle lattice, where high symmetry points $\Gamma$ (0, 0, 0), $M$(0, $\pi$, 0) and $K$ ($2\pi$/3, $2\pi$/3, 0) are marked. c, The schematic picture of the origin of band structure in monolayer (Bi$_{1-x}$Cr$_x$)$_2$Te$_3$/GdI$_2$, starting from the atomic orbitals of Bi and Gd, the following three steps are required to understand the band
structure: (I) the chemical bonding and crystal field effect, (II) the FM exchange coupling, and (III) SOC effect. E$_f$ is the Fermi level (dashed line).}\label{fig1}
\end{figure}

The basic mechanism for the QAH effect is the band inversion between spin polarized bands in magnetic TIs. In this Letter, we propose a new class of materials to realize the QAH effect. Since the highest $T_c$ in semiconductors is still far below room temperature, artificial materials are the only way. Layered crystals structures allow the production of monolayers and the synthesis of new compounds via combination of mono- and multilayers have been already demonstrated~\cite{Geimd2013}. By combining two topologically trivial FM materials together, the topological insulator property arises from the relative spin polarized band inversion~\cite{Liu2008a} between the two different FM materials of the thin films. In particular, based on first-principles calculations, for the interface between monolayer (Bi,Sb)$_2$Te$_3$ and GdI$_2$, we find that the spin polarized $p_z$-characterized conduction band from Cr-doped (Bi,Sb)$_2$Te$_3$ will invert with the spin polarized $d_{z^2}$-characterized valence band from Gd, resulting in a new QAH insulator with the  non-trivial band gap up to 38~meV. As we know, GdI$_2$ is a halfmetallic ferromagnet, where the ferromagnetism is contributed by the transition metal element Gd and its Curie temperature $T_c$ is around 300~K. A high $T_c$ is thus expected for monolayer GdI$_2$ and Cr$_x$(Bi,Sb)$_{2-x}$Te$_3$ interface. All these properties in this system are distinct from the previous proposals in magnetically doped TIs, such as Cr-doped Bi$_2$Se$_3$-family~\cite{Yu2010}, where a high $T_c$ and a large band gap are hard to achieve simultaneously.

Both Cr$_x$(Bi,Sb)$_{2-x}$Te$_3$ and GdI$_2$ are layered triangle lattice compounds interconnected along $c$-axis by van der Waals interactions, which makes their thin films chemically stable. Cr$_x$(Bi,Sb)$_{2-x}$Te$_3$ is a FM insulator with the in-plane lattice constant decreases from 4.35 {\AA} to 4.15 {\AA} as the content of Cr increases~\cite{Zhou2006}. Experimentally, GdI$_2$ crystallizes in the well-known 2H-MoS$_2$ structure (194 space group), in which each Gd layer are sandwiched by two layers of I atoms with the trigonal prismatic geometry~\cite{Kasten1984}. The in-plane lattice constant of GdI$_2$ is about 4.075 {\AA}, resulting in a lattice mismatch 2\% $\sim$ 6\% (depending on the content of Cr) with Cr$_x$(Bi,Sb)$_{2-x}$Te$_3$. At room temperature, GdI$_2$ show a metallic behavior and FM transition with saturation magnetic moment about 7.33~$\mu_{\mathrm{B}}$/Gd, but becomes insulating at low temperature (below 150~K)~\cite{Ahn2000}. Our calculations do find that the bulk of  GdI$_2$ is a bad semimetal. However, for monolayer, it becomes insulating with a band gap about 0.2~eV, where the dispersions originated from the interlayer interaction are eliminated due to quantum confinement.

There are lots of stacking configurations to combine Cr$_x$(Bi,Sb)$_{2-x}$Te$_3$ and GdI$_2$ together. Comparing to other configurations, such as A(Te)-B(Bi/Cr)-C(Te)-A(Bi/Cr)-B(Te)-A(I)-C(Gd)-A(I), A(Te)-B(Bi/Cr)-C(Te)-A(Bi/Cr)-B(Te)-B(I)-A(Gd)-B(I) and A(Te)-B(Bi/Cr)-C(Te)-A(Bi/Cr)-B(Te)-C(I)-B(Gd)-C(I), we predict that the film shown in Fig.~\ref{fig1}a, \emph{i.e.} A(Te)-B(Bi/Cr)-C(Te)-A(Bi/Cr)-B(Te)-C(I)-A(Gd)-C(I) is the most stable one. Although there are some energy difference between different configurations, the electronic band structures of them look very similar. In addition, the site of Cr in (Bi,Sb)$_2$Te$_3$ also have very weak influence on the band structures. Therefore, we will focus on the configuration shown in Fig.~\ref{fig1}a as representative in this paper.

We perform the first-principles density functional theory (DFT)~\cite{Hohenberg1964,Kohn1965} calculations and PAW potentials ~\cite{Blochl1994,Kresse1999} by the Vienna Ab-initio Simulation Package (VASP)~\cite{Kresse1993,Kresse1996}. Perdew-Burke-Ernzerhof-type ~\cite{Perdew1996} generalized gradient approximation + Hubbard U correction (DFT+U)~\cite{ Anisimov1991,Dudarev1998} with U = 6~eV (J = 0) on Gd's $4f$ orbitals and U = 3~eV (J = 1.5~eV) on Cr's $3d$ orbitals are used. The results are also double checked by HSE functional~\cite{Heyd2003}, which is known as more accurate description for the band structures of semiconductors. SOC effect is considered self-consistently in the calculations. The kinetic energy cut-off is fixed to 400~eV. 10$\times$10$\times$2/6$\times$6$\times$2 $k$-mesh are used for half doping and other Cr contents films respectively. For all the films, the vacuum region are thicker than 12 {\AA}, and the lattice constants, as well as the atomic positions are fully optimized with the accuracy smaller than 0.01 eV/{\AA}.

\begin{figure}[tbp]
\includegraphics[clip,scale=0.33,angle=270]{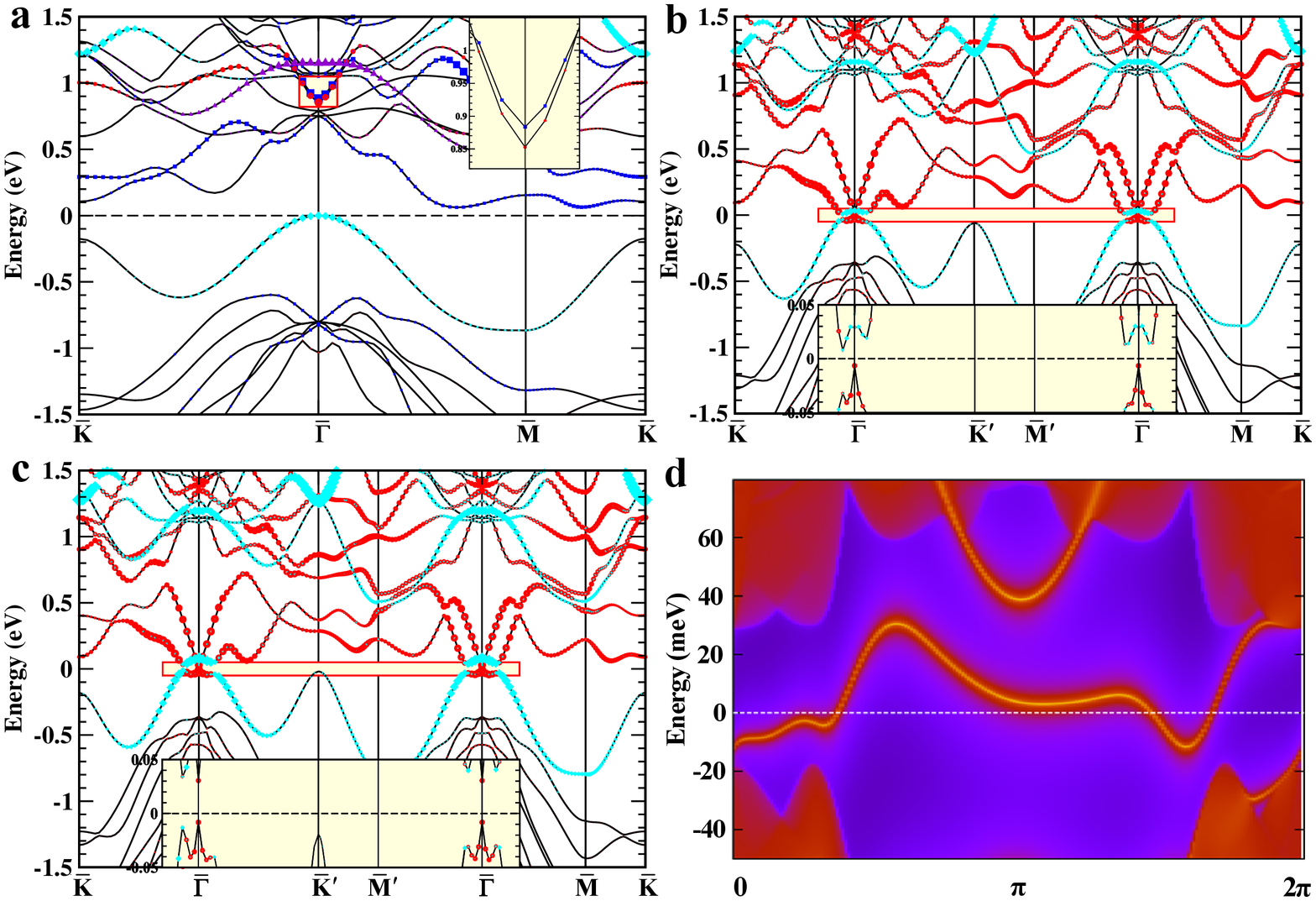}
\caption{(Color online) a, Spin-polarized band structures of BiCrTe$_3$/GdI$_2$ film without SOC,
in which red circles and blue squares denote the projections
to the down and up spin $p_z$ orbital of Bi, while the cyan diamonds and purple triangles mean the projections
to the up and down spin . The inset is the zoom-in of the red square region to show the magnetic splitting of $p_z$ orbital.
b and c, The band structures with SOC for $E=0$ (no gating electric field) and  $E=-0.1$~V/\AA case,
where the projections to Bi p orbitals and Gd $d_{z^2}$ orbital are indicated by red circles and cyan diamonds, respectively.
The insets in b and c are the zoom-in around the Fermi level to show the insulating gap clearly.
d, Calculated edge state of BiCrTe$_3$/GdI$_2$ based on the band structures shown in  Fig. 2c.
All Fermi levels are defined at 0 eV.}
\end{figure}

In  Fig.~\ref{fig1}c, we show a schematic picture of the origin of band inversion in monolayer Cr$_x$(Bi,Sb)$_{2-x}$Te$_3$ and GdI$_2$. Take the atomic energy levels of Bi ($6s^26p^3$), Gd ($4f^75d^16s^2$), Te($5s^25p^4$) and I($5s^25p^5$) as the starting point. At stage \uppercase\expandafter{\romannumeral1}, the chemical bonding and crystal field in Cr-doped (Bi,Sb)$_2$Te$_3$ and GdI$_2$ are considered separately: 1) In Cr-doped (Bi,Sb)$_2$Te$_3$, chemical bonding between Bi and Te push down all of the Te states, giving rise to the $p$-characterized valence bands of Te . On the other hand, the lifted up Bi $p$ orbitals are split into two degenerate manifolds $p_x, p_y$ and single $p_z$ state by crystal field. After this splitting, the conduction band closest to the Fermi level (E$_f$) is the $p_z$ state of Bi. 2) In GdI$_2$, chemical bonding push down all of the states of I far away from the Fermi level. Strong correlation effect split $f$ orbitals of Gd into up and down Hubbard bands, which are far away from Fermi level too. So we neglect $f$ states in this analysis. The trigonal prismatic coordination of the Gd atoms splits their $d$ orbitals into three groups, $d_{z^2}$, $d_{xy}, d_{x^2-y^2}$ and $d_{xz}, d_{yz}$. With this splitting, the remaining one $5d$ electron of Gd half occupies on $d_{z^2}$ state. At stage \uppercase\expandafter{\romannumeral2}, the $d_{z^2}$ band of Gd split into two branches due to the magnetic polarization, making the up spin branch fully occupied and GdI$_2$ insulating.  Because of the magnetic ordering of Cr, the $p_z$ conduction band of Bi is split in a way that down spin branch moves toward the Fermi level. Therefore, after the magnetic exchange interactions taking into account, the two bands closest to the Fermi level turn out to be $\left|d_{z^2},\uparrow\right\rangle$ (valence band) and $\left|p_z,\downarrow\right\rangle$ (conduction band) respectively. At the last stage \uppercase\expandafter{\romannumeral3}, after SOC effect are considered in, orbital and spin angular momenta mix together, while the total angular momentum is preserved. Therefore, $\left|p_z,\downarrow\right\rangle$ states becomes $\left|J_z^{\text{eff}} = -1/2\right\rangle$, which is very similar to the `split off band' in the Kane model~\cite{Voon2009}. Due to the huge SOC effect of Bi,  the energy of $\left|J_z^{\text{eff}} = -1/2\right\rangle$ will be dramatically dropped down and get inverted with $\left|J_z^{\text{eff}} = 1/2\right\rangle$ state, which originated from  $\left|d_{z^2},\uparrow\right\rangle$ of Gd. Because $d$ orbital and $p$ orbital have different parity, this kind of $d-p$ inversion at $\Gamma$ point usually leads to non-trivial Berry's phase for the occupied bands. Therefore, the system becomes topologically non-trivial Chern insulator if a full band gap exists, and its Chern number is determined by $\Delta J_z^{\text{eff}}$, \emph{i.e.} $C = 1$.

Based on the analysis of the band sequence discussed above, a simple $2\times2$ model, with $|d_{z^2},\uparrow\rangle$ and $|p_z,\downarrow\rangle$ as the basis, can be introduced to describe the spin polarized band inversion,
\begin{equation}
H_{\text{eff}} =\begin{pmatrix}
M & Dk_-\\
Dk_+ & -M
\end{pmatrix}
\end{equation}
where $k_\pm=k_x\pm ik_y$, and $M=M_0-B(k_x^2+k_y^2)$ is the mass term expanded to the second order, with
parameters $M_0>0$ and $B>0$ to ensure the band inversion. Since the two bases have opposite parity, the off-diagonal
element has to be odd in $k$. In addition, it has to have the form of $k_\pm$ to conserve the angular momentum along the $z$
direction.

In order to justify this effective model, we have performed the first principles calculations on the (Bi$_{1-x}$Cr$_x$)$_2$Te$_3$/GdI$_2$ films. First, we focus on the half doping case, and show its spin-polarized band structures (without SOC) in  Fig. 2a. Our calculations confirm that the magnetic moments on Cr and Gd prefer ferromagnetic arrangement with 0.33 meV lower than anti-ferromagnetic configuration. The calculated moments are 3.20 $\mu_{\mathrm{B}}$/Cr and 7.48 $\mu_{\mathrm{B}}$/Gd, similar to previous calculations~\cite{Yu2010,Zhang2013} and measurements~\cite{Ahn2000}. As shown in  Fig. 2a, one single-layer BiCrTe$_3$ is a normal insulator due to the confinement effect, in which the conduction bands from Bi and Cr are well separated with the valance band from Te by a gap about 0.6 eV. Within this gap lies an occupied spin-up band of Gd, making BiCrTe$_3$/GdI$_2$ film as a narrow gap insulator if SOC is not considered.

\begin{figure}[tbp]
\includegraphics[clip,scale=0.33,angle=270]{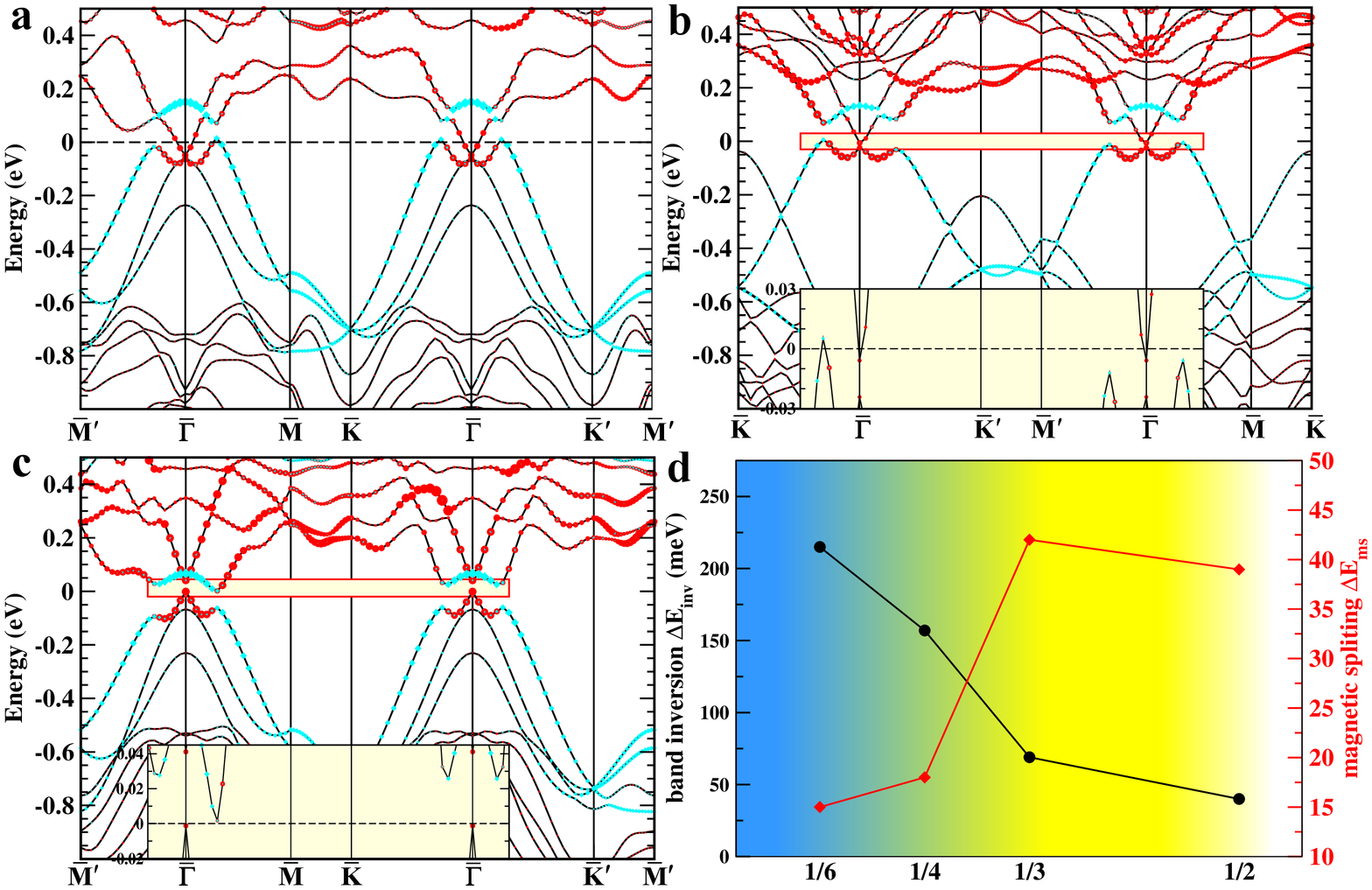}
\caption{(Color online) a, b and c show the band structures with SOC for 1/6, 1/4 and 1/3 Cr-doped Bi$_2$Te$_3$ and GdI$_2$ films.
The size of red circles and cyan diamonds mean the projections to Bi p orbitals and Gd $d_{z^2}$ orbital respectively.
The insets in b and c are the zoom-in of the red square region around the Fermi level. All Fermi levels are defined at 0 eV.
d, The schematic phase diagram of Cr-doped Bi$_2$Te$_3$ and GdI$_2$ films,
in which the yellow region on right side means the Chern insulator, the blue region on left side indicates the topological metal.
Among of them, the transition region means the tunable topological semi-metal.
  }
\end{figure}

Next, we carry fully relativistic calculations, to study the influence of SOC on the electronic structures by aligning the magnetic moments perpendicular to the film, since the direction of magnetic moment in Cr-doped (Bi,Sb)$_2$Te$_3$ is confirmed to be along $c$-axis~\cite{Chang2013,Zhou2006}. In the presence of SOC, $\left|d_{z^2},\uparrow\right\rangle$ of Gd becomes $\left|J_z^{\text{eff}} = 1/2\right\rangle$. Meanwhile, $p_z$ orbitals evolve into two 'spin hole' like bands by reconstruction with $p_x$ and $p_y$ orbitals. Because inversion symmetry is broken in this system, these two 'spin hole' like bands show large Rashba splitting, resulting in one inner band and one outer band with W-shape around $\bar{\Gamma}$ point. The outer and inner bands are further split at $\bar{\Gamma}$ point due to magnetic exchange interactions. Approximately, the inner band is called $\left|J_z^{\text{eff}} = 1/2\right\rangle$,  since it contains $\left|p_z,\uparrow\right\rangle$ component at $\bar{\Gamma}$ point; the outer band, which originated from $\left|p_z,\downarrow\right\rangle$,  is  defined as $\left|J_z^{\text{eff}} = -1/2\right\rangle$; the energy difference between the inner band ($\left|J_z^{\text{eff}} = 1/2\right\rangle$) and outer band ($\left|J_z^{\text{eff}} = -1/2\right\rangle$) at $\bar{\Gamma}$  point is defined as $\Delta E_{\mathrm{ms}}$. Our calculations suggest  $\Delta E_{\mathrm{ms}}=39$~meV at Cr half doping case. Comparing with  Fig. 2a, huge SOC effect of Bi drops down the  'spin hole' like bands dramatically. As shown in  Fig. 2b, even though there is still a normal gap bigger than 0.5 eV between Bi's $p$ orbitals and Te's $p$ orbitals, the $\left|J_z^{\text{eff}} = -1/2\right\rangle$ band from Bi is inverted with $\left|J_z^{\text{eff}} = 1/2\right\rangle$ band from Gd at $\bar{\Gamma}$ point, and re-open an insulating gap about 15 meV. According to our previous analysis, this kind of $d-p$ band inversion can give rise to a topological non-trivial Chern insulator. We note that, due to both time reversal symmetry and inversion symmetry are broken in ferromagnetic BiCrTe$_3$/GdI$_2$ film, the band structures along $\bar{\Gamma}$-$\bar{K}$  and $\bar{\Gamma}$-$\bar{K'}$  directions show different behavior in  Fig. 2b and  Fig. 2c.

Another advantage of the freestanding films is that one can modulate the band inversion continuously by an external gating electric field E~\cite{Kou2013}. In  Fig. 2c, we show the well-modulated band structures of BiCrTe$_3$/GdI$_2$ film by $E=-0.1$~V/\AA, in which the optimal topological insulating gap 38 meV is achieved. In order to confirm the system's topological properties, we carry out the calculations of edge states by constructing the Green functions~\cite{Sancho1985} for the semi-infinite edge based on Maximally Localized Wannier functions method~\cite{Marzari1997,Souza2001}. The results based on the band structures with optimal gap ( Fig. 2c) are shown in  Fig. 2d, in which one topologically protected chiral edge state connecting $\left|J_z^{\text{eff}} = 1/2\right\rangle$ and $\left|J_z^{\text{eff}} = -1/2\right\rangle$ presents clearly, consistent with our previous analysis that $C = 1$.

We also perform calculations on systems with other Cr contents, such as 1/6, 1/4, 1/3 doping, trying to study the topological properties as a function of Cr doping. The results are shown in  Fig. 3. Our calculations suggest that, at 1/3 doping case, the system is also Chern insulator with a small gap about 3.6~meV (see Fig. 3c), in which the magnetic splitting at $\bar{\Gamma}$ point $\Delta E_{\mathrm{ms}}$ is about 44 meV,  a little enhanced  than half doping case. However, when Cr content keeps decreasing, the magnetic splitting $\Delta E_{\mathrm{ms}}$ becomes smaller and smaller (see  Fig. 3d), while the band inversion between $\left|J_z^{\text{eff}} = 1/2\right\rangle$ from Gd and  $\left|J_z^{\text{eff}} = -1/2\right\rangle$ from Bi becomes deeper and deeper. Therefore both the cases of 1/4 and 1/6 doping are topologically non-trivial semimetal, in which both valence band and conduction band are partially occupied at different momentum $k$ (see  Fig.~3a and  Fig.~3b). We summarize the topological properties evolution with Cr content in  Fig.~3d, in which $\Delta E_{\mathrm{inv}}$ is defined by  the energy level of $\left|J_z^{\text{eff}} = 1/2\right\rangle$ from Gd minus the energy level of $\left|J_z^{\text{eff}} = -1/2\right\rangle$ from Bi. As shown in  Fig. 3d, one can find that: 1) Band inversion $\Delta E_{\mathrm{inv}}$ increases monotonically with the decrease of Cr concentration, which is because of the `split off' bands dropping deeper and deeper due to the increased SOC effect~\cite{Zhang2013a}. 2) Magnetic splitting $\Delta E_{\mathrm{ms}}$ is mainly decreasing when the Cr concentration becomes smaller. This is consistent with the fact that Curie temperature $T_c$ decreases with Cr concentration reduction in experiment~\cite{Zhou2006}. As a result, when $x>1/3$, \emph{i.e.} the yellow region on right side in  Fig.~3d, (Bi$_{1-x}$Cr$_x$)$_2$Te$_3$/GdI$_2$ films are the Chern insulator, in which QAH effect can be realized. In contrast, with the weekly doping $x<1/6$, \emph{i.e.} the blue region on left side in  Fig.~3d, the system is topological metal with $d-p$ band inversion. Between them, the transition region indicates the tunable topological semi-metal, which means one can modulate this semi-metal to a Chern insulator by applying experimentally reasonable gate voltage or by Sb doping. Take 1/4 doping as an example, by applying electric gating field $E=0.2$~V/\AA, a Chern insulator with a gap about 18~meV can be achieved.

\begin{figure}[tbp]
\includegraphics[clip,scale=0.285]{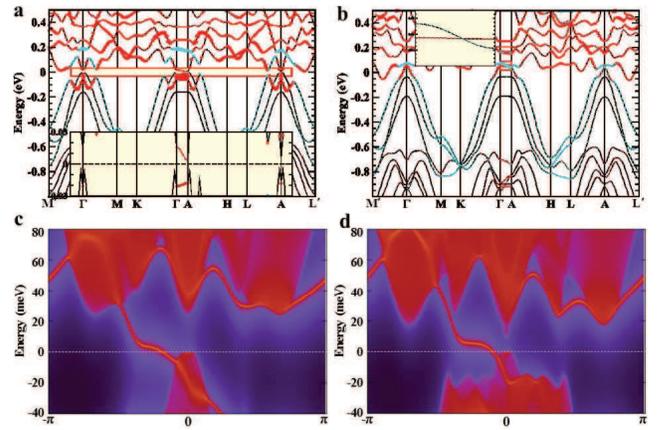}
\caption{(Color online) a and b show the band structures with SOC for (Bi$_{2/3}$Cr$_{1/3} $)$_2$Te$_3$ and (Sb$_{2/3}$Cr$_{1/3} $)$_2$Te$_3$ superlattice respectively.
The size of red circles and cyan diamonds mean the projections to Bi p orbitals and Gd $d_{z^2}$ orbital respectively.
The insets in a is the zoom-in of the red square region around the Fermi level. The insets in b is the zoom-in of the red square region to show Weyl node.
c and d is the d calculated edge state of (Bi$_{2/3}$Cr$_{1/3} $)$_2$Te$_3$ superlattice for $k_z = 0$ and $k_z = \pi$ plane respectively. The Fermi levels are defined at 0 eV.  }
\end{figure}

Finally we would like to address the topological properties of the superlattice. Because Cr-doped (Bi,Sb)$_2$Te$_3$ and GdI$_2$ are coupled by the van der Waals interactions along z-direction, the dispersion of the supperlattice along $k_z$ is very weak. If the $d-p$ band inversion is big enough, making all $k_z$-plane's Chern number equaling 1, such system can be called 3D Chern insulator~\cite{Xu2011}. Here we show the calculated band structures on (Bi$_{2/3}$Cr$_{1/3} $)$_2$Te$_3$ /GdI$_2$ supperlattice in  Fig. 4a, where the  $\left|J_z^{\text{eff}} = -1/2\right\rangle$ state from Bi is always lower than the $\left|J_z^{\text{eff}} = 1/2\right\rangle$ state from Gd at all $k_z$ and a full insulating gap about 8 meV present. This indicates that 3D Chern insulator is realized in (Bi$_{2/3}$Cr$_{1/3} $)$_2$Te$_3$ /GdI$_2$ supperlattic, which can be confirmed by the edge state calculations for $k_z = 0$ and $k_z = \pi$ plane (shown in Fig.~4c and 4d). On the other hand, if we tune $\Delta E_{\mathrm{inv}}$ to a suitable size, making that the $d-p$ band inversion only happens at $\Gamma$ point while $A$ point ($k_z = \pi$) remains a normal insulating gap, the system will be a nontrivial semimetal with topologically unavoidable Weyl nodes~\cite{Xu2011} located at the phase boundary separating $C = 1$ and $C = 0$ planes. In Fig. 4b, we show the numerical results on (Sb$_{2/3}$Cr$_{1/3} $)$_2$Te$_3$ /GdI$_2$ supperlattice, confirming that such topologically unavoidable Weyl node form in it, even though the Weyl nodes do not cross the Fermi level due to some $d$ bands from Cr are dropping below E$_f$ in (Sb$_{2/3}$Cr$_{1/3} $)$_2$Te$_3$.

This work is supported by the US Department of Energy, Office of Basic Energy Sciences, Division of Materials Sciences and Engineering under Contract No.~DE-AC02-76SF00515, and by the Defense Advanced Research Projects Agency Microsystems Technology Office, MesoDynamic Architecture Program (MESO) through contract numbers N66001-11-1-4105, and  partly by FAME, one of six centers of STARnet, a Semiconductor Research Corporation program sponsored by MARCO and DARPA. G.X. would like to thank for the support from 973 program of China (No.2013CB921704) and NSF of China.

\end{document}